# Electro-mechanical control of an optical emitter using graphene

Antoine Reserbat-Plantey[1,*], Kevin G. Schädler[1,*], Louis Gaudreau[1,*], Gabriele Navickaite[1], Johannes Güttinger[1], Darrick Chang[1], Costanza Toninelli[2], Adrian Bachtold[1], and Frank H.L. Koppens[1]

[1]ICFO-Institut de Ciencies Fotoniques, Mediterranean Technology Park, 08860 Castelldefels, Barcelona, Spain  [2]LENS and CNR-INO, Via Nello Carrara 1, 50019 Sesto Fiorentino, Italy.

[*]:*These authors contributed equally*
*Corresponding authors: adrian.bachtold@icfo.es, frank.koppens@icfo.es*

**Active, in situ control of light at the nanoscale remains a challenge in modern physics and in nanophotonics in particular[1–3]. A promising approach is to take advantage of the technological maturity of nano-electromechanical systems (NEMS) and to combine it with on-chip optics[4–6]. However, in scaling down the dimensions of such integrated devices, the coupling of a NEMS to optical fields becomes challenging. Despite recent progress in nano-optomechanical coupling[7–9], active control of optical fields at the nanoscale has not been achieved with an on-chip NEMS thus far. Here, we show a new type of hybrid system, which consists of an on-chip graphene NEMS suspended a few tens of nanometers above nitrogen-vacancy centres (NVC), which are stable single photon emitters embedded in nano-diamonds. Electromechanical control of the photons emitted by the NVC is provided by electrostatic tuning of the graphene NEMS position, which is transduced to a modulation of NVC emission intensity. The optomechanical coupling between the graphene displacement and the NVC emission is based on near-field dipole-dipole interaction. This class of optomechanical coupling increases strongly for smaller distances, making it suitable for devices with nanoscale dimensions. These achievements hold promise for the selective control of single-emitter arrays on chip, optical spectroscopy of individual nano-objects, integrated optomechanical information processing and quantum optomechanics.**

Recent work has shown that graphene is an ideal platform for both nanophotonics[10–14] and nano-mechanics[15–17]. Here, we demonstrate a single device, combining these two platforms (Fig. 1). In this device, the transduction between nano-motion and an optical field is due to a strong modification of an emitter's relaxation rate and light emission when graphene is placed in its near field[13,18–23], at nanometer-scale distances. The coupling strength increases strongly for shorter distances, and is enhanced due to graphene's 2D character and linear dispersion. As such, this near-field hybrid optomechanical coupling mechanism between graphene and a point dipole is intrinsically nanoscale in comparison to the evanescent coupling involving micron-scale cavities and waveguides in previous works[4,24–26]. In addition, thanks to its electro-mechanical properties, graphene NEMS can be actuated and deflected electrostatically over few tens of nanometers with modest voltages applied to a gate electrodes[16,27]. The graphene motion can thus be used to modulate the light emission, while the emitted field can be used as a universal probe[18,19,22] of the graphene position.

The coupling between an emitter and graphene can manifest itself as various ways, such as Stark shift[28], dipolar-coupling induced modification of the emission intensity[19–21,29] or energy (Casimir-Polder[30]), or as energy transfer to graphene plasmons[10,13]. In our experiment, the graphene-induced emission intensity modification is particularly suitable for coupling the graphene displacement and the NVC emission. This effect is due to non-radiative energy transfer (n-RET) and is mediated by dipolar interactions between the emitting point dipole and induced dipoles in graphene, as shown schematically in Fig. 1-a. As a consequence, it gives rise to a diverging decay rate[18] $\Gamma_g \propto d^{-4}$ of the emitter in the presence of graphene at a separation $d = $ 5-50 nm. Therefore, the emission is reduced with decreasing graphene-emitter separation (cf. SI). Here, graphene offers the advantage to be a 2D gapless broadband energy sink. First, this enables energy transfer rates described by material-free parameters, thus defining the concept of a universal nano-ruler[18,19,22]. Second, the enhanced dipolar coupling strength and stronger distance dependence ($d^{-4}$ compared to $d^{-3}$ for bulk materials[19]) makes the near-field dipolar interaction a more effective divergent coupling mechanism between a graphene NEMS and a fluorescent emitter.

To harness near-field dipolar interactions for nano-optomechanical coupling, we propose and demonstrate a novel type of integrated hybrid device as shown in Fig. 1-b. Our device consists of a graphene membrane designed to be suspended some 10-50 nm above a fluorescent nitrogen vacancy centre[31] embedded in a nano-diamond, as shown in Fig. 1-cd. By applying a combination of a DC and AC voltage to the conducting graphene membrane relative to the doped silicon backgate, we can simultaneously drive the resonator at radio frequencies and also control the graphene-NVC separation. Thus, our device employs the NVC as a transducer of the resonator's nano-motion which modulates the emission intensity, and complementary enables electro-mechanical control of the emitter.

We fabricate large arrays of hybrid devices (10 to 100 devices per chip), and we address them individually. The emitted light is monitored by a custom-made low-temperature scanning confocal microscope (Fig.1-f), simultaneously recording maps of the reflectance (Fig.1-e) from the device. Reflection measurements allow us to detect the nano-motion of the graphene resonators by interferometry (cf. SI), yielding *e.g.* a map of the mechanical resonance frequency $f_m$ as shown in Fig. 1-g. Colocalization of the emitters and graphene resonators is revealed by the optical measurements shown in Figs. 1f and 1g.

In order to quantitatively study and control the near-field interaction between the NVC and the graphene resonator, we tune the graphene-emitter separation electrostatically. The membrane is attracted towards the NVC by applying a potential difference $V_g^{dc}$ between the backgate and the graphene drum. Optical interferometry measurements (cf. SI) show that the static deflection of the graphene scales as $\xi_{static} \propto \left(V_g^{dc}\right)^2$, in agreement with electrostatic actuation. These measurements allow actuation calibration on the order of 1.2 ± 0.1 nm/V² for the sample shown here. For a given value of $V_g^{dc}$, the graphene-emitter separation is $d_{G-NVC}\left(V_g^{dc}\right) = d_0 - \xi_{static}\left(V_g^{dc}\right)$, where $d_0$ is the initial graphene-emitter separation extracted from the measured device topology (see Fig.1-c). Such electrostatic actuation provides *in situ* and stable control of the graphene deflection from its initial position to the point of contact with the nano-diamond.

We use this *in-situ* control to experimentally verify that the graphene-emitter coupling is due to n-RET by measuring the NVC emission as a function of the membrane position. As shown in Fig. 2, we observe a non-linear reduction of the NVC emission as the membrane is electrostatically deflected towards the nano-diamond. This quenching behaviour can be explained by n-RET when the graphene approaches an emitter. As introduced above, the emitter decay rate $\Gamma_g(d)$ has non-linear dependence on the separation and thus induces an emission reduction (black line $\phi_g \propto \left(\Gamma_g(d)\right)^{-1}$. We remark that the observations cannot be a result of an interferometric modulation of the excitation intensity due to graphene deflection. Namely, for our device geometry (oxide thickness and hole depth) the interferometric effect leads to an increase of the emission upon decreasing $d_{G-NVC}$, in contrast to the measurements shown in Fig. 2 (cf. SI for further discussion).

We exploit the concept of the universal distance ruler, coined in previous studies[18,19,22], to extract the separation between the graphene and the localised emitter, a quantity that is difficult to extract using far-field or local probe techniques. To this end, we fit the data with a model (free of graphene material parameters, hence universal) where we consider energy transfer from a single NVC to graphene[18]. Further, we consider the background signal shown in Fig. 2 at small separations to originate from NVC's embedded deeper within the nano-diamond, at distances where their interaction with graphene is negligible (cf. SI). Good agreement of our data with the n-RET model shows that emission measurements provide an indirect optical probe of the graphene position.

Additionally, our hybrid device enables high frequency and local control of individual emitters at sub-wavelength scales. To demonstrate this concept, we show optomechanical transduction of radio-frequency graphene resonator nano-motion to NVC emission, by performing time-resolved emission measurements. During a mechanical oscillation cycle, the graphene-emitter separation $d_{G-NVC}$ periodically varies with an amplitude $\delta z$ and the graphene position is imprinted on the NVC decay rate $\Gamma_g(z_0 + \delta z)$. To observe this, we drive the resonator capacitively at frequency $f_{drive}$ and simultaneously perform time-correlated single photon counting of the emitted photons over a few mechanical periods. By repeating such synchronized acquisition, we obtain a histogram of photon arrival times modulated at $f_{drive}$, as shown in Fig. 3-a. As such, this transduction mechanism involves three successive steps: i) the initial electro-mechanical actuation of the membrane, followed by ii) a quasi-instantaneous optomechanical transduction due to n-RET ($\frac{c}{d_{G-NVC}} \gg f_m$, c being the speed of light), and finally iii) a conversion into a time-resolved electronic signal at the single photon counting module. Additional interferometric measurements enable calibration of the driven oscillation amplitude to be approximately 1 nm at resonance (cf. SI). This implies that the optomechanical transduction step is linear as $\delta z \ll d_{G-NVC}$, resulting in the observed sinusoidal emission modulation despite the non-linear dependence of emission on $d_{G-NVC}$. Our hybrid device allows for the first time to imprint the nano-motion of a graphene resonator onto the emitted field of an emitter acting as a local probe, which is a key feature for applications in the context of quantum photonics integrated circuits (QPIC)[32–34].

In order to reveal the mechanical spectrum of the graphene resonator in the NVC emission, we extract the modulation depth $A_{FFT}(f_{drive})$ defined as the Fourier component of the emission time-traces at different frequencies $f_{drive}$. At mechanical resonance, both the amplitude of motion $\delta z(f_m)$ and $A_{FFT}(f_m)$ are greatest. Indeed, by sweeping $f_{drive}$ through $f_m$ (independently measured by interferometry), we can reconstruct the mechanical spectrum of the graphene resonator (Fig. 3-b) through the near-field interaction modulation between the graphene resonator and a localized emitter.

The magnitude of this oscillatory component in the emission can be controlled actively, thus demonstrating *in-situ* control of the transduction strength. In order to observe such control, we record $A_{FFT}(f_m)$ while varying the stationary separation $d_{G-NVC}$, as shown in Fig. 3-c. Here, the differential emission $\Delta A_{FFT} = A_{FFT}(f_m)/A_r(f_m)$ is the measured emission modulation amplitude $A_{FFT}(f_m)$, normalised to the resonant oscillation amplitude $A_r(f_m)$ as obtained from interferometry. This normalisation is necessary to compensate the increase of $A_r$ with increasing backgate voltage as $\delta z(f_{drive}) \propto \chi_m V_g^{dc} V_g^{ac}(f_{drive})$, where $\chi_m$ is the mechanical susceptibility. The measurements shown in Fig. 3-c reveal that $\Delta A_{FFT}$ diminishes with increasing $V_g^{dc}$. Indeed, while the near-field interaction diverges with decreasing $d_{G-NVC}$, the observed emission and thus the transduced signal is quenched. Our data can be fitted by the derivative of NVC emission with respect to $d_{G-NVC}$, as expected from the non-linear dissipative coupling model introduced previously in Fig. 2. We find that the largest transduction would be obtained for a separation of 35±3 nm. These results, summarized in Fig 3-c, show that we achieve active control of the optomechanical coupling strength by tuning the separation between a local emitter and a vibrating graphene NEMS.

Finally, we explore the localized nature of the near-field coupling mechanism in the normal plane through time-resolved emission from our hybrid device over the whole area of the driven resonator. A spatial map of $A_{FFT}(f_m)$ (inset of Fig. 3-b) shows a stronger signal at the NVC site, as expected for such n-

RET coupling between graphene and a point-like emitter. This observation highlights the intrinsic localization of the interaction, confined within a sub-wavelength volume. As such, the atomically small emitter can be used as a local transducer of the graphene NEMS motion, which enable eigenmode shape reconstruction. On the other hand, driving the NEMS at higher-order mechanical modes, with sub-wavelength spatial modulations, allows addressing and coupling to individual emitters distributed over separations unresolvable with far-field optics.

In conclusion, we have realised a novel device comprising a graphene NEMS coupled to nitrogen vacancy centers by near-field dipolar coupling. Our work offers interesting perspectives for lock-in detection of weak fluorescence signals, NEMS position sensing, electro-mechanical control of emitters on chip and fast electro-mechanical light modulation at single photon level. Based on the system presented, we envision a similar device which harnesses vacuum forces, inducing a divergent and dispersive optomechanical coupling between lifetime-limited linewidth quantum emitters and the motion of 2-dimensional systems[30]. Such emblematic implementation of the interaction of a two-level system with a phonon bath, would enable coherent manipulation of mechanical and optical degrees of freedom at the level of single quanta[7,35,36].

**Acknowledgments**

The authors thank O. Arcizet, J. Bertelot, V. Bouchiat, F. Dubin, J. Moser, C. Muschik, M. Lewenstein, M. Lundeberg, J. Osmond, K. Tielrooij, I. Tsioutsios and P. Weber.
KGS acknowldeges funding from Europhotonics. LG acknowledges financial support from Marie-Curie International Fellowship COFUND and ICFOnest program. AB acknowledges support by the ERC starting grant 279278 (CarbonNEMS), and support by the EE Graphene Flagship (contract no. 604391). FK acknowledges support by the Fundacio Cellex Barcelona, the ERC Career integration grant 294056 (GRANOP), the ERC starting grant 307806 (CarbonLight), and support by the EE Graphene Flagship (contract no. 604391) and project GRASP (FP7-ICT-2013-613024-GRASP).

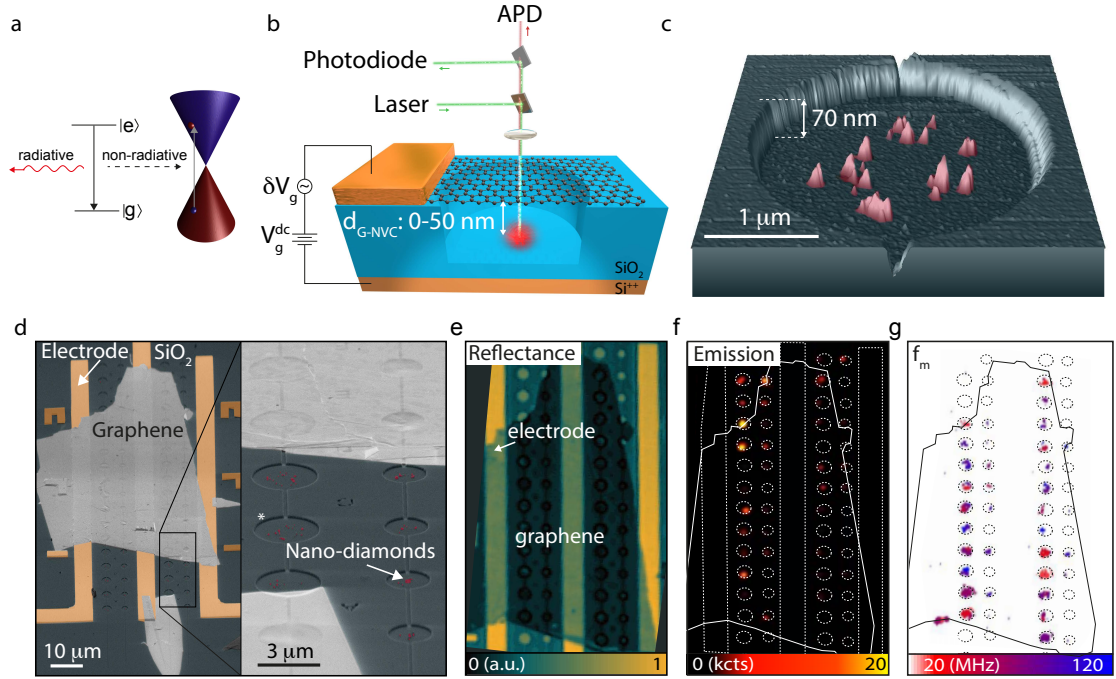

**Figure 1 : Graphene-NVC hybrid optomechanical device. a**: Energy diagrams of an optical emitter and graphene at the K point of the Brillouin zone (Dirac cone). For small separations $d_{G-NVC} < 50$ nm, the relaxation of the excited emitter is predominantly due to near-field dipole-dipole interaction by excitation of electron-hole pairs in graphene. **b**: Sketch of the hybrid optomechanical device. The graphene resonator is driven and displaced electrostatically by DC and AC voltages $V_g^{dc}$ and $\delta V_g$, while its nano-motion is measured optically *via* the emitter using single photon counters (APD) and by interferometry. **c**: False colour AFM topology of nano-diamonds (red) deposited in the centre of a hole etched into $SiO_2$. **d**: False colour scanning electronic micrograph (SEM) of arrays of hybrid graphene devices. The labelled hole corresponds to **c**. Graphene is closely suspended over the nano-diamonds (0-50 nm) and clamped at the edges of the holes by Van der Waals interactions. **e**: False colour confocal reflection map of the same device shown in **c** at T=3K. At each position, both emission (**f**) and a mechanical spectrum is recorded. The mechanical resonance frequency $f_m$ is shown in (**g**).

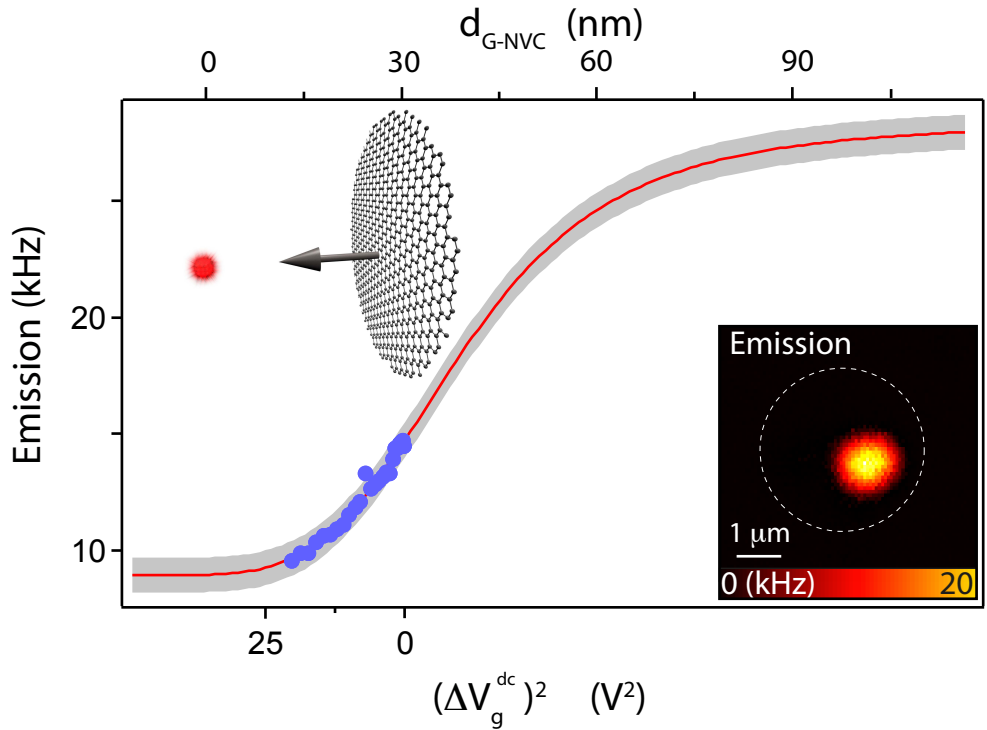

**Figure 2 : In-situ control of the dipolar coupling between emitter and graphene drum.** Dependence of measured NVC emission intensity (blue) on the graphene-emitter separation $d_{G-NVC}$, controlled by electrostatic deflection of the membrane. With increasing $\left(\Delta V_g^{dc}\right)^2$, the membrane approaches the NVC, thereby reducing $d_{G-NVC}$ and quenching the NVC emission. Data can be fitted with a n-RET model (red), which also allows deflection calibration. Inset: optical emission map of the hybrid system showing localised NVC emission (dashed line: graphene resonator outline).

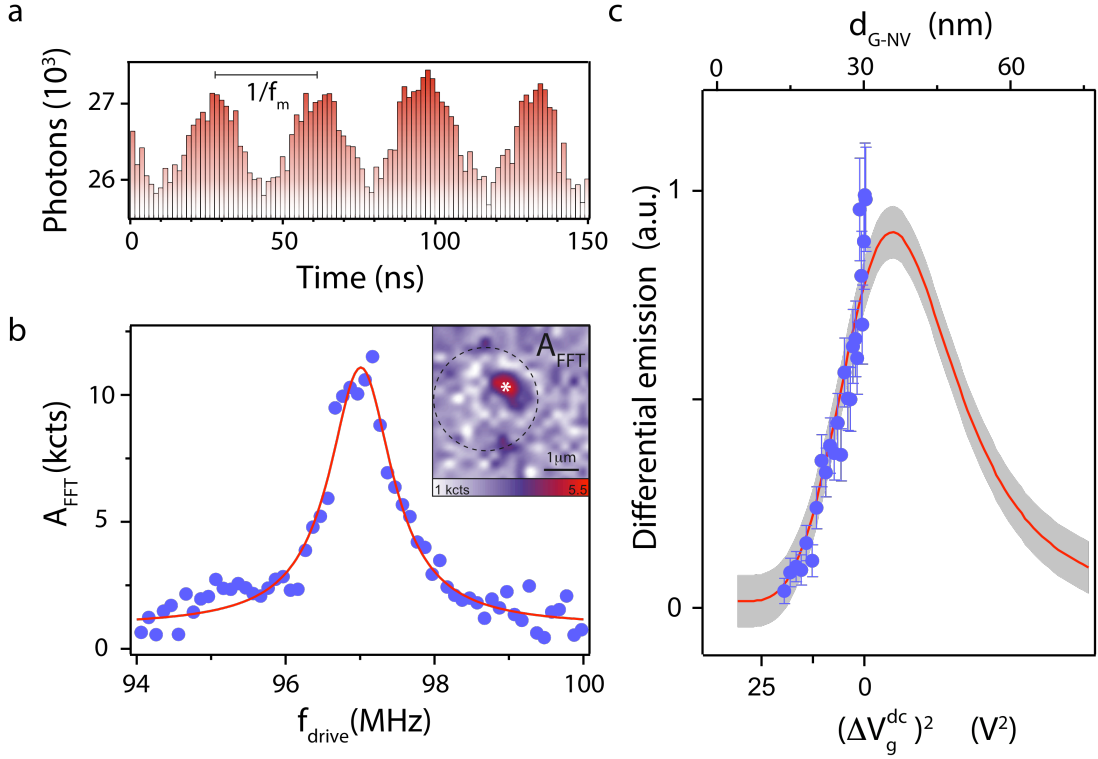

**Figure 3 : Time-resolved readout of graphene resonator nano-motion via NVC emission. a**: Time trace of NVC emission (bars) modulated by a driven graphene membrane oscillating at $f_m$ in its near field. Distance-dependent dipolar emitter-graphene coupling imprints the nano-motion of the graphene membrane onto the emission. **b**: Mechanical resonance obtained from differential oscillating NVC emission. Each point corresponds to the amplitude of the Fourier component at $f = f_{drive}$ of the emission time trace (as in **a**). A Lorentzian fit of the data (red line) yields the same value for $f_m$ as obtained independently by optical interferometry. Inset: spatial map of Fourier component at $f = f_{drive}$, localised around the emitter position (*). **c**: Dependence of measured (blue) and extrapolated (red) differential NVC emission intensity on the graphene-emitter separation $d_{G-NVC}$, normalised by the membrane's resonant oscillation amplitude $A_m$ (which increases with $V_g$, thus reducing the measurement uncertainty for small separations $d_{G-NV}$).